\documentclass[iop,apj,appendixfloats]{emulateapj}
\usepackage{apjfonts}
\usepackage{graphicx}
\usepackage{amsmath}
\usepackage{amssymb}
\usepackage{color}

\gdef\xx[#1]{\textcolor{red}{#1}}

\gdef\kms{km\,s$^{-1}$}
\gdef\msun{M$_{\odot}$}

\gdef\gal{NGC1052-DF4}
\gdef\dffour{NGC1052-DF4}
\gdef\dftwo{NGC1052-DF2}
\gdef\group{NGC\,1052}

\newcommand{\GG}[1]{}

\lefthead{van Dokkum et al.}
\righthead{A second galaxy missing dark matter}
\slugcomment{Accepted for publication in ApJ Letters}
\begin{document}

\newcommand\XXX[1]{{\textcolor{red}{\textbf{x\ #1\ x}}}}

%\title{
%A Distance of $18.7\pm 1.7$\,Mpc for the Dark Matter Deficient Galaxy
%\blob\ from an Analysis of the Color-Magnitude Diagram and a Megamase-TRGB-SBF Distance Ladder}
%Calibrated to the H$_2$O Megamaser Galaxy NGC4258

\title{A second galaxy missing dark matter in the NGC\,1052 group}

\author{Pieter van Dokkum\altaffilmark{1}, Shany Danieli\altaffilmark{1},
Roberto Abraham\altaffilmark{2}, Charlie Conroy\altaffilmark{3},
%J.~M.\ Diederik Kruijssen\altaffilmark{4}, Allison Merritt\altaffilmark{5},
Aaron J.\ Romanowsky\altaffilmark{4,5}
%Shany Danieli\altaffilmark{1},
%Yotam Cohen\altaffilmark{1},
%Aaron 
%Charlie Conroy\altaffilmark{2}
\vspace{8pt}}

\altaffiltext{1}
{Astronomy Department, Yale University, 52 Hillhouse Ave,
New Haven, CT 06511, USA}
%\altaffiltext{2}
%{University of California Observatories, 1156 High Street, Santa
%Cruz, CA 95064, USA}
%\altaffiltext{3}
%{Department of Physics and Astronomy, San Jos\'e State University,
%San Jose, CA 95192, USA}
%\altaffiltext{2}
\altaffiltext{2}
{Department of Astronomy \& Astrophysics, University of Toronto,
50 St.\ George Street, Toronto, ON M5S 3H4, Canada}
\altaffiltext{3} 
{Harvard-Smithsonian Center for Astrophysics, 60 Garden Street,
Cambridge, MA, USA}
%\altaffiltext{4} 
%{Astronomisches Rechen-Institut, Zentrum f\"ur Astronomie der Universit\"at
%Heidelberg, M\"onchhofstra{\ss}e 12-14, D-69120 Heidelberg, Germany}
%\altaffiltext{5}
%{Max-Planck-Institut f\"ur Astronomie, K\"onigstuhl 17, D-69117
%Heidelberg, Germany}
\altaffiltext{4}
{Department of Physics and Astronomy, San Jos\'e State University,
San Jose, CA 95192, USA}
\altaffiltext{5}
{University of California Observatories, 1156 High Street, Santa
Cruz, CA 95064, USA}

\begin{abstract}

The ultra-diffuse galaxy NGC1052-DF2 has a 
very low velocity dispersion, indicating that it
has little or no dark matter. Here we report the discovery of a second
galaxy in this class, residing in the same group. \gal\ closely
resembles NGC1052-DF2 in terms of its size, surface brightness,
and morphology; has a similar
distance of $D_{\rm sbf}=19.9\pm 2.8$\,Mpc;
and also has a population
of luminous globular clusters
extending out to $\geq 7$\,kpc from the center of the galaxy.
Accurate radial
velocities of the diffuse galaxy light and seven of the globular
clusters were obtained with the Low Resolution Imaging Spectrograph
on the Keck\,I telescope. The velocity of the diffuse
light is identical to the median velocity of the clusters,
$v_{\rm sys}=\langle v_{\rm gc}\rangle=1445$\,\kms,  and
close to the central velocity of the NGC1052 group.
% and the velocity of NGC1052
%itself (1510\,\kms).
The  rms spread of the globular cluster velocities is very small at
$\sigma_{\rm obs}=5.8$\,\kms. Taking observational uncertainties
into account we determine an
intrinsic velocity dispersion of
$\sigma_{\rm intr}=4.2^{+4.4}_{-2.2}$\,\kms, consistent with the
expected value from the stars alone ($\sigma_{\rm stars}
\approx 7$\,\kms) and lower than expected from a standard NFW
halo ($\sigma_{\rm halo}\sim 30$\,\kms).
We conclude that NGC1052-DF2 is not an isolated
case but that a class of such
objects exists. The origin of these large, faint galaxies
with an excess of luminous globular clusters and an apparent lack of
dark matter is, at present, not understood.

\end{abstract}

\keywords{
galaxies: evolution --- galaxies: structure}

%\facility{Keck:I (LRIS)}

\section{Introduction}

Sensitive surveys using state of the art telescopes
have identified large numbers of intrinsically-large
galaxies with very low
surface brightness (e.g., {van Dokkum} {et~al.} 2015; {Koda} {et~al.} 2015; {van der Burg}, {Muzzin}, \&  {Hoekstra} 2016). 
These ``ultra diffuse galaxies'' (UDGs), with sizes $R_{\rm e}>
1.5$\,kpc and central surface brightness $\mu_g>24$\,mag\,arcsec$^{-2}$,
have been found in many different
environments (including the Local Group; {Martin} {et~al.} 2016; {Torrealba} {et~al.} 2019),
and have a wide range of properties (see, e.g., {Merritt} {et~al.} 2016).

One of the most intriguing UDGs that have been studied so far is
\dftwo\ in the \group\ group.
Using a combination of Hubble Space Telescope {\em HST}
Advanced Camera for Surveys (ACS) imaging and 
Keck spectroscopy, we determined that this galaxy has
an unusual population of
luminous globular cluster-like objects ({van Dokkum} {et~al.} 2018c).
Furthermore, from the radial velocities of ten of these globular clusters 
we determined that the galaxy appears to have little or
no dark matter ($\lesssim 10^8$\,\msun; {van Dokkum} {et~al.} 2018b; {Wasserman} {et~al.} 2018). Both aspects are surprising:
the globular cluster luminosity function was thought to be
universal ({Rejkuba} 2012), and a galaxy with a stellar mass
of $\sim 2\times 10^8$\,\msun\ should have a dark matter
mass of $\sim 6\times 10^{10}$\,\msun\ ({Behroozi}, {Wechsler}, \&  {Conroy} 2013b).

Although these unexpected results were initially greeted with some
skepticism (e.g., {Martin} {et~al.} 2018; {Laporte}, {Agnello}, \&  {Navarro} 2018; {Nusser} 2018; {Hayashi} \& {Inoue} 2018;
{Ogiya} 2018; Trujillo et al.\ 2019),
recent studies have corroborated the unusual nature of \dftwo:
the distance to the galaxy
was placed on surer footing
($D=19-20$\,Mpc; {van Dokkum} {et~al.} 2018a; {Blakeslee} \& {Cantiello} 2018) and, crucially,
the low mass of \dftwo\ has been confirmed by measurements of the stellar velocity
dispersion (Emsellem et al.\ 2019; Danieli et al.\ 2019). Specifically,
{Danieli} {et~al.} (2019) find $\sigma=8.5^{+2.3}_{-3.1}$\,\kms\ from
a high resolution spectrum taken with the Keck Cosmic Web Imager.
%, similar to
%that of Local Group galaxies that are a factor of $\sim 100$ less
%luminous than \dftwo\
%(we note that
%a competing claim is made in {Emsellem} {et~al.} 2018, based on data
%of much lower spectral resolution).

At this point, the central unanswered question is whether \dftwo\
is an isolated case or representative of a population of similar
galaxies. This is important for judging the likelihood
of interpretations that require unusual
orbits or viewing angles (see, e.g., {Ogiya} 2018) and, most
importantly, for judging the relevance of \dftwo\ for our ideas
about galaxy formation and the relation between dark matter and normal
matter. With the important exception of tidal dwarfs
({Bournaud} {et~al.} 2007; {Gentile} {et~al.} 2007; {Lelli} {et~al.} 2015), it is
often thought
that a gravitationally-dominant dark matter halo is the {\em sine qua non}
for the formation of a galaxy (White \& Rees 1978).
If galaxies such as \dftwo\ are fairly common we may have to
revise our concept of what a galaxy {\em is}, and come up with 
alternative pathways for creating such large and
relatively massive stellar systems.
%\footnote{Tidal dwarfs are examples of such systems.}

Here we report the discovery of a galaxy that shares essentially all of
\dftwo's unusual properties, to a remarkable degree. It is
in the same group, has a similar size, luminosity, and color,
the same morphology, the same population
of luminous globular clusters, and the same extremely low
velocity dispersion.

\begin{figure*}[htbp]
  \begin{center}
  \includegraphics[width=0.9\linewidth]{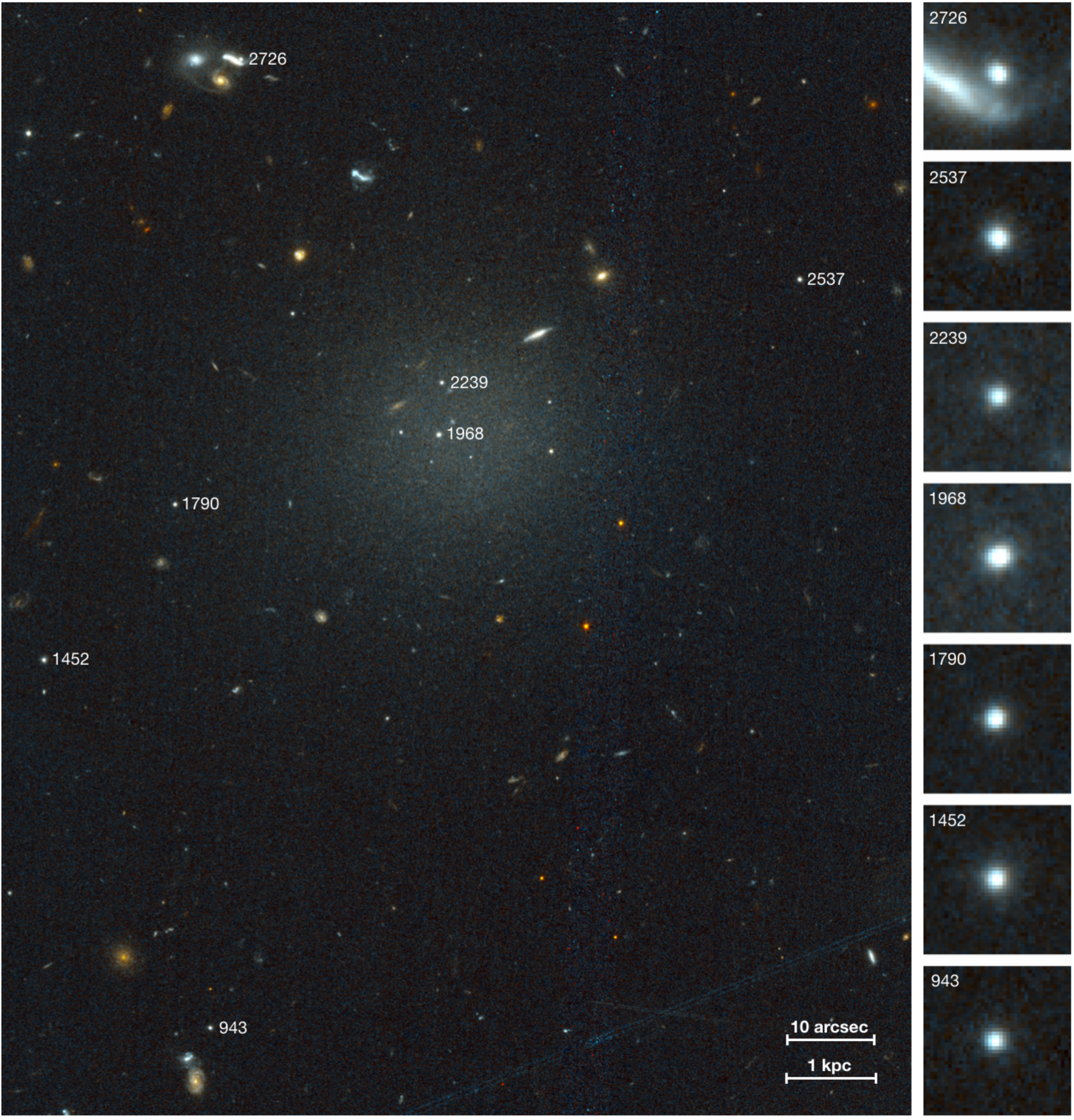}
  \end{center}
\vspace{-0.2cm}
    \caption{{\em HST} ACS image of \dffour, created from the $V_{606}$ and
$I_{814}$ bands. The galaxy has a smooth, spheroidal morphology with
a low S\'ersic index ($n=0.79$). The highlighted objects are
spectroscopically-confirmed globular clusters.
}
\label{figure1.fig}
\end{figure*}

\section{\dffour}

\dffour\ is a low surface brightness galaxy in the field of the elliptical
galaxy NGC\,1052. It is
part of a sample of 23
objects that we identified in images
taken with the
Dragonfly Telephoto Array ({Abraham} \& {van Dokkum} 2014) and followed up with
the ACS on {\em HST}
(see {Cohen} {et~al.} 2018).
Unlike \dftwo, which had been described earlier by
{Fosbury} {et~al.} (1978) and {Karachentsev} {et~al.} (2000), \dffour\ was
discovered with Dragonfly.\footnote{This is a somewhat academic
point as the galaxy is not particularly faint; it is
clearly visible in Plate 1 of {Fosbury} {et~al.} (1978) and in many
other imaging datasets.}
The {\em HST} image is shown in Fig.\ \ref{figure1.fig}.

Basic parameters, as derived from the {\em HST} imaging, are given in
{Cohen} {et~al.} (2018).
Its surface brightness fluctuation distance is $D_{\rm sbf}=19.9\pm 2.8$\,Mpc,
and as highlighted in Fig.\ 5 of {van Dokkum} {et~al.} (2018a)
its color-magnitude diagram is very similar to that of \dftwo.
We infer that the galaxy is part of the \group\ group, and use
$D=20$\,Mpc for distance-dependent quantities.
The galaxy is well-fitted by a 2D {Sersic} (1968) profile with a
S\'ersic index of $n=0.79$, an axis ratio of $b/a=0.89$, central
surface brightness $\mu(V_{606},0)=23.7$, and major
axis half-light radius $R_{\rm e}=1.6$\,kpc. These properties
place the galaxy just inside the UDG selection box (see Fig.\ 10
of {Cohen} {et~al.} 2018). The total absolute magnitude is
$M_{V,606}=-15.0\pm 0.1$, corresponding to $L_{V,606}=(7.7\pm 0.8)\times
10^7$\,L$_{\odot}$. Assuming $(M/L)_{{\rm stars},V}=(2.0\pm
0.5)\,{\rm M}_{\odot}/{\rm L}_{\odot}$ (see {van Dokkum} {et~al.} 2018b) the
implied total stellar mass is $M_{\rm stars} = (1.5\pm
0.4)\times 10^8$\,\msun.

At first glance \dffour\ seemed to lack the
spectacular population of bright globular cluster-like objects that
initially
drew our attention to \dftwo\ (see {van Dokkum} {et~al.} 2018b, 2018c).
However, careful inspection
of the {\em HST} imaging data shows that the \dffour\ field
actually {\em does} contain objects with similar sizes, colors, and
apparent magnitudes as the globular clusters in \dftwo\ -- but that they
are even more spread out relative to the body of the galaxy.
This unexpected finding inspired us to obtain spectra of
these candidate globular clusters, to determine whether they are
actually in a single structure and, if so, to use them to constrain
the mass of \dffour.

\noindent
\begin{deluxetable*}{lcccccccc}
\tablecaption{NGC1052-DF4 Globular Clusters\label{globs.table}}
\tablehead{\colhead{Id} & \colhead{RA} & \colhead{Dec}  & $R$\tablenotemark{a}
& $M_{V,606}$ & $V-I$ & $v$ & $r_{\rm h}$ & $\epsilon$ \\
 & (J2000)  & (J2000) & [kpc]\tablenotemark{b} & [mag]\tablenotemark{b} & [mag]\tablenotemark{c} & [km\,s$^{-1}$]
 & [pc] & }
\startdata
NGC1052-DF4 & 2$^{\rm h}39^{\rm m}15.11^{\rm s}$ & $-8\arcdeg6\arcmin58\farcs6$
& \nodata & $-15.0$ & $0.32$ & $1444.6_{-7.7}^{+7.8}$ & $1600$ & 0.11 \\
\hline
GC-2726 & 2$^{\rm h}39^{\rm m}16.75^{\rm s}$ & $-8\arcdeg6\arcmin16\farcs7$
&4.69 & $-9.2$ & $0.38$ & $1441.2_{-4.8}^{+4.9}$ & $4.9\pm 0.6$ & $0.14\pm 0.04$ \\
GC-2537 & 2$^{\rm h}39^{\rm m}12.53^{\rm s}$ & $-8\arcdeg6\arcmin41\farcs4$
&4.08 & $-9.2$ & $0.36$ & $1451.0_{-3.3}^{+3.6}$ & $4.1 \pm 0.5$ & $0.08 \pm 0.06$ \\
GC-2239 & 2$^{\rm h}39^{\rm m}15.23^{\rm s}$ & $-8\arcdeg6\arcmin53\farcs0$
&0.57 & $-8.6$ & $0.34$ & $1457.1_{-5.5}^{+4.6}$ & $5.4\pm 0.5$ & $0.16\pm 0.04$ \\
GC-1968 & 2$^{\rm h}39^{\rm m}15.25^{\rm s}$ & $-8\arcdeg6\arcmin58\farcs8$
&0.20 & $-9.8$ & $0.27$ & $1445.4_{-2.3}^{+2.6}$ & $3.4\pm 0.3$ & $0.36 \pm 0.05$ \\
GC-1790 & 2$^{\rm h}39^{\rm m}17.24^{\rm s}$ & $-8\arcdeg7\arcmin06\farcs7$
&3.17 & $-9.0$ & $0.31$ & $1438.4_{-4.6}^{+4.8}$ & $3.2 \pm 0.4$ & $0.19\pm 0.08$ \\
GC-1452 & 2$^{\rm h}39^{\rm m}18.23^{\rm s}$ & $-8\arcdeg7\arcmin24\farcs1$
&5.13 & $-9.1$ & $0.32$ & $1445.5_{-4.1}^{+4.0}$ & $5.1\pm 0.6$ & $0.04\pm 0.03$ \\
GC-943 & 2$^{\rm h}39^{\rm m}16.98^{\rm s}$ & $-8\arcdeg8\arcmin5\farcs3$
&7.01 & $-8.6$ & $0.34$ & $1445.1_{-5.2}^{+5.0}$ & $3.3\pm 1.1$ & $0.41\pm 0.17$ \\
\hline
BG-2844 & 2$^{\rm h}39^{\rm m}20.37^{\rm s}$ & $-8\arcdeg6\arcmin18\farcs6$
&\nodata & \nodata & \nodata& $z=0.2298$ & \nodata & \nodata \\
BG-254 & 2$^{\rm h}39^{\rm m}11.55^{\rm s}$ & $-8\arcdeg9\arcmin03\farcs0$
&\nodata & \nodata & \nodata& $z=0.2557$ & \nodata & \nodata
\enddata
\tablenotetext{a}{Distance from the center of the galaxy.}
\tablenotetext{b}{For an assumed distance of $D=20$\,Mpc.}
\tablenotetext{c}{$V_{606}-I_{814}$ from HST/ACS, in the AB system.}
\end{deluxetable*}

\begin{figure}[htbp]
  \begin{center}
  \includegraphics[width=1.0\linewidth]{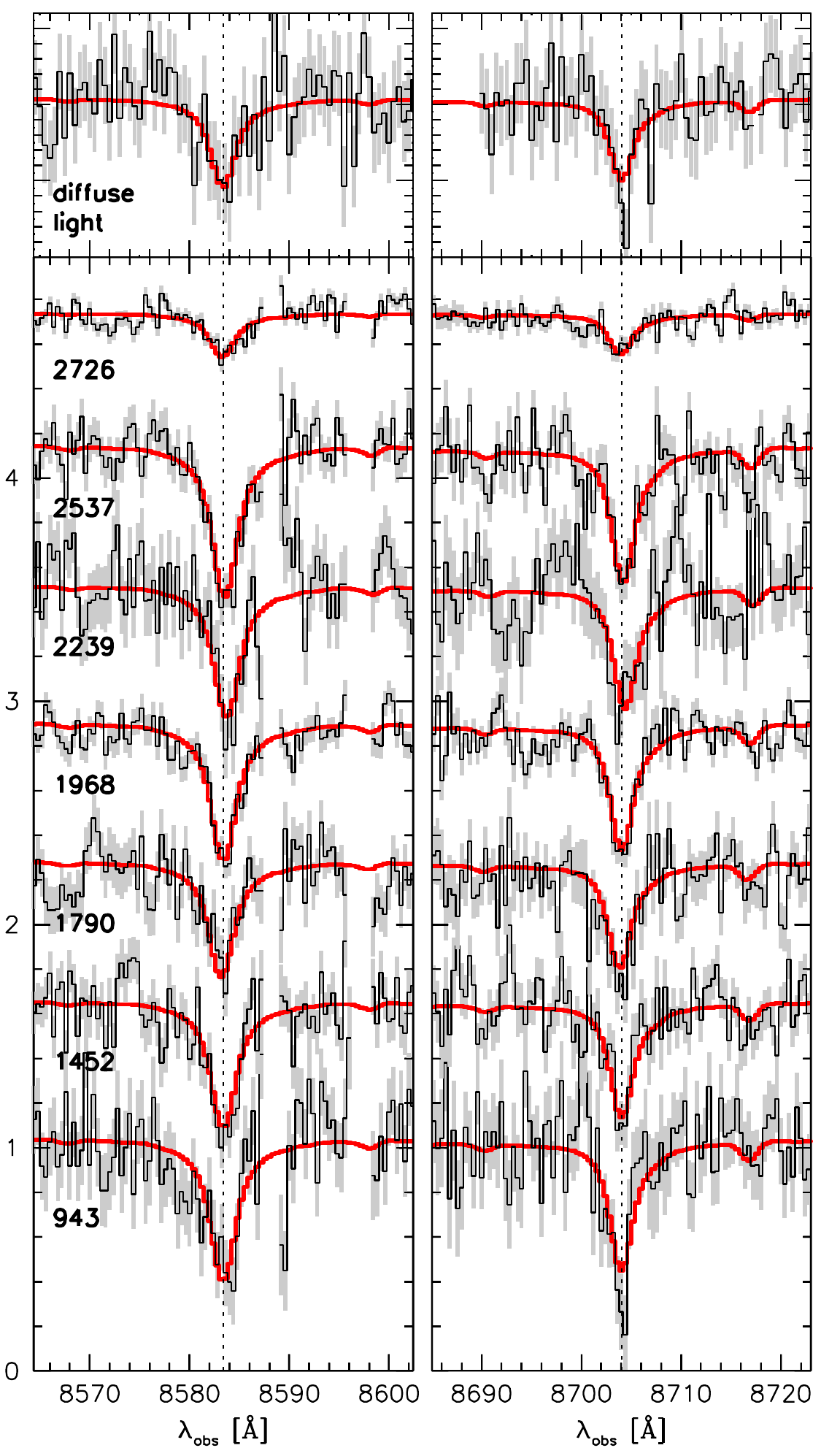}
  \end{center}
\vspace{-0.2cm}
    \caption{Keck/LRIS spectra of the diffuse light of \dffour\ and
    seven bright globular cluster-like objects (offset for clarity),
in the region of the strongest Ca triplet lines. The uncertainties are shown in grey,
and the 
model spectra that were fitted to the data to determine  radial velocities
are in red. The vertical dashed line indicates the median
velocity of $\langle v\rangle=1445$\,\kms.
%The spread around this velocity
%is extremely small.
}
\label{specs.fig}
\end{figure}

\section{Spectroscopy}

We observed compact objects in the \dffour\ field with the
dual arm
Low Resolution Imaging Spectrograph (LRIS) on the Keck\,I telescope.
The sample selection is modeled upon the properties of
the confirmed clusters in \dftwo. We used SExtractor ({Bertin} \& {Arnouts} 1996)
to measure total magnitudes, colors, and FWHM sizes of objects in
the {\em HST} images, following the methodology outlined in
{van Dokkum} {et~al.} (2018c). Priority was given to objects with
$I_{814}<23$, $0.20<V_{606}-I_{814}<0.43$, and $0 \farcs 12<{\rm FWHM}<
0\farcs 30$, as all 11 clusters from {van Dokkum} {et~al.} (2018c)
satisfy these criteria.
In \dffour, seven objects fall within these limits. They have a mean
total magnitude of $\langle I_{814}\rangle =22.10$ with a $1\sigma$
rms spread of 0.39\,mag. Objects just outside these selection limits
were given lower priority.\footnote{There are three objects
with $23.0<I_{814}<23.5$ that satisfy the color and size criteria.}

All seven high priority
objects could be fitted in a single multi-slit mask, along
with four lower priority targets. This mask was observed 
on November 6 2018 for a total of 19,800\,s, split over eleven 1,800\,s
exposures. Conditions were excellent. On the blue side the
low resolution
300\,lines\,mm$^{-1}$ grism blazed at 5000\,\AA\ was used, and on the
red side the 1200\,lines\,mm$^{-1}$ grating blazed at 9000\,\AA.
Here only the high resolution
red-side observations are discussed.
The slit width of the mask is $0\farcs 8$, providing a resolution of
$\sigma_{\rm instr}\approx 25$\,\kms\ near the Calcium triplet.

\begin{figure*}[htbp]
  \begin{center}
  \includegraphics[width=0.9\linewidth]{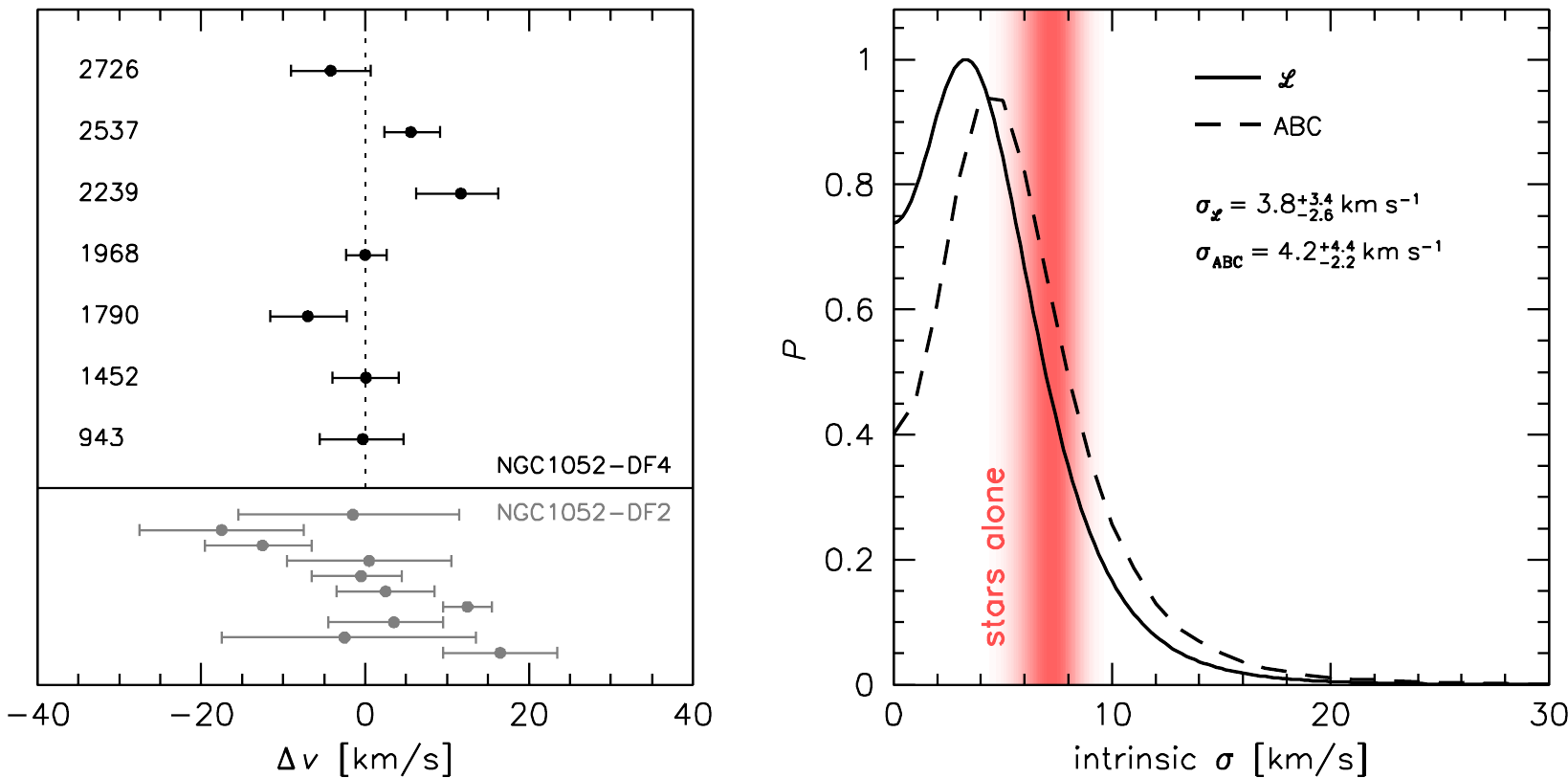}
  \end{center}
\vspace{-0.2cm}
    \caption{{\em Left panel:} Graphical representation of the
globular cluster velocities in \dffour\ (top) and
\dftwo\ (bottom), relative to the median (which is very close to the
systemic velocity for both galaxies). {\em Right panel:}
Constraints on the intrinsic velocity dispersion of
\dffour\ from the seven clusters, using the likelihood
estimator (solid) and approximate bayesian computation (dashed).
The expected velocity dispersion for a normal
dark matter halo is $\sim 30$\,\kms.
The  expected dispersion from the stellar mass alone, with no
dark matter, is $\approx 7$\,\kms.
}
\label{result.fig}
\end{figure*}

The data reduction follows the same procedures as those outlined in
{van Dokkum} {et~al.} (2018b). Briefly, the slit edges and the sky lines are used
to correct the spectra for distortions. The sky line modeling and
subtraction is done using the method
of {Kelson} (2003), which minimizes
interpolation-related residuals. The sky lines are also used for
wavelength calibration. Each individual 1,800\,s exposure
is analyzed independently to limit the effects of flexure on the
distortion modeling and wavelength calibration. The exposures
are combined using optimal weighting, and 1D spectra are
extracted by weighting each line in the 2D spectrum
by the S/N ratio.

An inspection of the spectra shows that all seven bright
globular cluster candidates indeed have strong absorption features
at the approximate redshift of the \group\ group. Two of the four lower priority
objects
turn out to be compact background galaxies and the remaining two
are too faint for a redshift measurement.
We show the spectra of the seven high priority targets
in Fig.\ \ref{specs.fig},
focusing on the regions near the
redshifted $\lambda \lambda 8542.09,8662.14$\,\AA\ lines of
the Ca triplet.\footnote{Note that the
spectrum of GC-2726 is blended with a star forming background
galaxy.}
The weaker $\lambda 8498.02$\,\AA\ line is mostly masked, as it
coincides with a strong sky line at the redshift of \dffour.
Despite a shorter total integration time
the spectra are of higher quality than those of 
most of the clusters in \dftwo\ ({van Dokkum} {et~al.} 2018b), due to 
better seeing and photometric conditions during the observations. The median
S/N ratio near
the Ca triplet is 17\,\AA$^{-1}$.

We also extract the spectrum of the diffuse galaxy light, in the following way.
The central two clusters (GC-1968 and GC-2239) were observed through a relatively long slit that
covers most of the extent of \dffour. We create an average sky spectrum from the other
nine slits and subtract this from the background spectrum of the central slit, after scaling. The
resulting spectrum is shown in the top panel of Fig.\ \ref{specs.fig}. The Ca triplet
lines are clearly detected.

\section{Kinematics}

\subsection{Radial Velocities}

Radial velocities of the diffuse light and
the seven clusters  are determined by fitting
the data with a synthetic
11\,Gyr, [Fe/H]=$-1$ stellar population synthesis model
({Conroy}, {Gunn}, \& {White} 2009; {Choi} {et~al.} 2016), convolved to the instrumental resolution.
The fit is performed
in the redshifted Calcium triplet region 8520\,\AA$<\lambda<8740$\,\AA, using
the {\tt emcee} MCMC algorithm ({Foreman-Mackey} {et~al.} 2013). The fit is regularized
by dividing the data and template by a polynomial of order
$100^{-1}\Delta \lambda$\,[\AA]. Two free parameters are fitted:
the velocity and an additive continuum offset, to account for any template
mismatch. With the exception of the blended spectrum of GC-2726 this
parameter is typically close to zero. Errors are determined
from simulations. In each simulation
the residuals of the fits are randomly shuffled
and the velocity fit is repeated. Residuals from
sky lines are shuffled separately from the rest of the spectrum.
The width of the
distribution of the resulting velocities is taken as the uncertainty
in the fit (see {van Dokkum} {et~al.} 2018b). The results are not sensitive
to the details of these procedures; owing to the high S/N ratio
of the spectra the velocities and the associated uncertainties
are very stable.\footnote{We note that we do not apply any corrections for possible
slit alignment errors; inspection
of alignment check images showed no evidence for offsets, but we cannot exclude
systematic errors at the level of a few km/s.}

The velocity of the diffuse light is $v_{\rm sys}=1444.6^{+7.8}_{-7.7}$\,\kms.
This is very close
to the average velocity of other galaxies in the \group\ group.
%Searching around the elliptical galaxy \group\ and 
Including \dftwo\
there are 22 galaxies in the NASA Extragalactic Database 
within a radius of two degrees centered on \group\ in the velocity range
$0<v<2500$\,\kms.
All are in the range
1241\,\kms\,$<v<$1805\,\kms, with a biweight mean of
$v_{\rm group}=1438 \pm 25$\,\kms\ and width $\sigma_{\rm group}=
128\pm 19$\,\kms.
The velocities of the globular clusters
are listed in Table 1 and displayed in the left panel
of Fig.\ \ref{result.fig}. The random uncertainties are small; the mean
$\pm 1\sigma$ error is $\pm 4$\,\kms. The median (mean) velocity
is $\langle v\rangle = 1445$\,\kms\ (1446\,\kms), identical to the systemic
velocity of the galaxy. This confirms that the clusters are associated with the
galaxy, and suggests that the globular cluster system is at rest with respect to the stars.
%\footnote{The
%present communis opinio doctorum is, perhaps, that \dftwo\ is a member
%of the NGC\,1052 group; the properties of \dffour\ should
%assuage any lingering concerns.}
%The similarity of the systemic velocity of
%\dffour\ and $v_{\rm group}$
%further strengthens the case that the galaxy
%is a member of the \group\ group at a distance
%of $\approx 20$\,Mpc.

\subsection{Velocity Dispersion}

The velocity range of the seven clusters is extremely small, echoing our
earlier result for \dftwo. The observed rms, before correcting for
observational uncertainties, is only 5.8\,\kms\ (compared to 10\,\kms\
for \dftwo).
%Furthermore, the uncertainties in the velocities
%are smaller than for
%\dftwo, due to the higher S/N ratio of the spectra (see
%left panel of Fig.\ \ref{result.fig}).
%As a result, despite the fact that we only have seven tracers
%the constraints on the intrinsic dispersion are actually stronger
%than for \dftwo.
%
We use two methods to determine the intrinsic dispersion $\sigma_{\rm intr}$
and its associated uncertainties. Both methods use a generative model
whose parameters are constrained by assessing the probability of
measuring the observed velocity distribution. The model is a simple
Gaussian with the center and width as free parameters.
The classical method, used extensively for determining
the kinematics of dwarf galaxies in the Local Group from
the velocities of individual stars
(e.g., {Martin} {et~al.} 2007), is
to construct the likelihood function:
\begin{equation}
\mathcal{L} = \prod_{i=1}^{i\leq 7}
\frac{1}{\sqrt{2\pi}\sigma_{\rm eff}}
\exp\left[-0.5\left( \frac{v_i-\mu}{\sigma_{\rm eff}}\right)^2\right],
\end{equation}
with $v_i$ the velocities of the individual tracers,
$\mu$ the mean of the
model, and $\sigma_{\rm eff}^2 = \sigma_{\rm intr}^2 + e_i^2$
with $\sigma_{\rm intr}$ the model dispersion
and $e_i$ the uncertainty in velocity $v_i$
(calculated by averaging
the positive and negative error bars). The likelihood,
marginalized over $\mu$, is
shown by the solid line in the right panel of Fig.\ \ref{result.fig}.
The likelihood analysis gives $\sigma_{\rm intr}=
3.8^{+3.4}_{-2.6}$\,\kms, where the uncertainties contain 68\,\%
of the probability. The 90\,\% (95\,\%) confidence
level upper limit is 8.6\,\kms\ (10.4\,\kms).

The second method is approximate Bayesian computation
(ABC; {Tavare} {et~al.} 1997), which
approximates the posterior distribution with simulations. For each set of
model parameters ($\mu$, $\sigma_{\rm intr}$) a large number of
simulated datasets $\hat{D} = (\hat{v}_1, \hat{v}_2, \ldots, \hat{v}_7)$
are created by randomly drawing velocities
from the model and linearly perturbing these values with errors that
are themselves randomly drawn from Gaussians of width ($e_1, e_2, \ldots,
e_7$). These $N$ datasets are compared to the
actual data $D$  through a summary statistic $S$. Simulations
that satisfy the criterion
\begin{equation}
\delta(S(\hat{D}),S(D)) < e
\end{equation}
are retained, with $\delta$ the absolute distance between the
summary statistics and
$e$ a small positive number.
For a given choice
of ($\mu$, $\sigma_{\rm intr}$)
the posterior is proportional to
the number of simulations that are retained.
ABC
does not assume a functional form of the likelihood,
and summary statistics can be chosen that are best suited to particular
situations (see, e.g., {van Dokkum} {et~al.} 2018b). Another
advantage is that it does not suffer from the ``small sample bias''
discussed in {Laporte} {et~al.} (2018).
%A disadvantage is that
%the association of individual data points with their individual
%errors is not retained.
We calculate the ABC posterior using $N=10^4$, $e=0.1$, and
the rms as the summary statistic (Fig.\ \ref{result.fig}).
ABC gives a similar result as the likelihood: $\sigma_{\rm intr}=
4.2^{+4.4}_{-2.2}$\,\kms. The small difference may reflect the
likelihood's sensitivity to small sample bias (see {Laporte} {et~al.} 2018).

\subsection{Implied Mass}

Quantitative constraints on the halo mass
are highly uncertain with seven tracers and require extensive modeling
(see, e.g., {Laporte} {et~al.} 2018; {Wasserman} {et~al.} 2018). Here we simply test the hypothesis
that there is no dark matter halo and all the mass is in the form of
stars.
Following {Beasley} {et~al.} (2016) and {van Dokkum} {et~al.} (2018b)
we estimate the mass within the outermost globular cluster using the
tracer mass estimator (TME) method of {Watkins}, {Evans}, \& {An} (2010):
\begin{equation}
M_{\rm TME}=\frac{C}{G}\langle(\Delta v')^2r^{\alpha}\rangle{}r_{\rm out}^{1-\alpha}.
\end{equation}
Here $\Delta v'= f^{-1}(v-1445.7)$ are the velocities of the individual
tracers, with $f=\sigma_{\rm obs}/\sigma_{\rm
intr}=1.4_{-0.7}^{+1.5}$. The parameter $\alpha$ is the slope of the potential
%(with density $\rho\propto r^{-(\alpha+2)}$) 
and $C$ is a constant given by
\begin{equation}
C=\frac{4\Gamma\left(\frac{\alpha}{2}+\frac{5}{2}\right)}{
\sqrt{\pi}\Gamma\left(\frac{\alpha}{2}+1\right)}
\,\frac{\alpha+\gamma-2\beta}{\alpha+3-\beta(\alpha+2)},
\end{equation}
with $\gamma$ the power-law slope of the 3D
density profile of the tracers and
$\beta=1-\sigma_{\rm t}^2/\sigma^2_{\rm r}$ the Binney anisotropy parameter.
For simplicity we assume that
$\beta=0$ and that the globular clusters trace the potential,
so that $\gamma=\alpha+2$.
If all the mass is in stars the potential is similar to that of a point mass
for most of the globular clusters; hence we use $\alpha=1$ and $\gamma=3$.

\begin{figure*}[htbp]
  \begin{center}
  \includegraphics[width=0.9\linewidth]{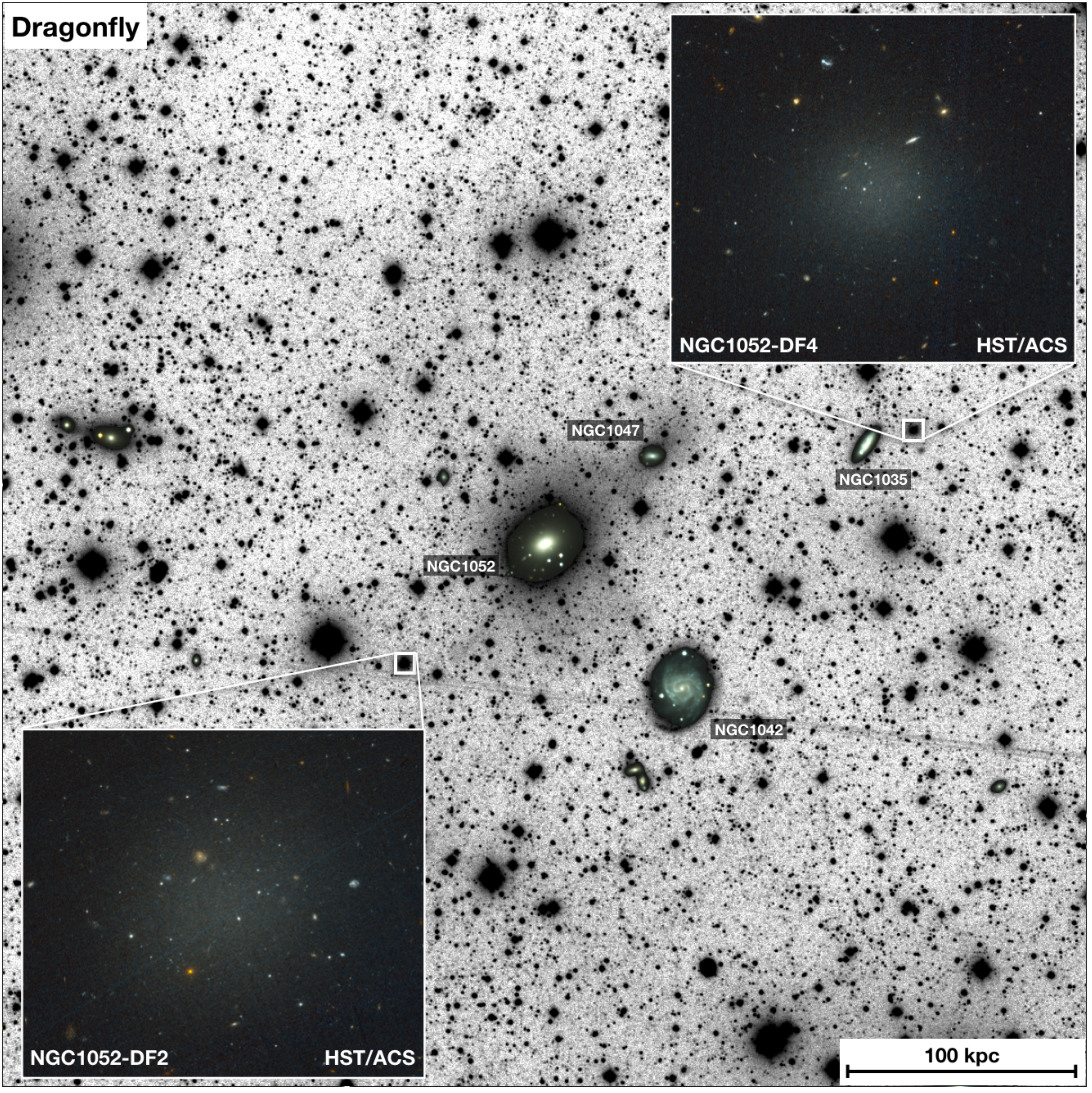}
  \end{center}
\vspace{-0.2cm}
    \caption{Central area of the summed
$g+r$ Dragonfly image of the \group\ field.
The displayed area covers $1.33\arcdeg \times 1.33\arcdeg$, corresponding
to $466$\,kpc$\,\times 466$\,kpc for a distance of 20 Mpc.
\dftwo\ and
\dffour\ are highlighted. The {\em HST} images span $93\arcsec \times 80\arcsec$
(9.0\,kpc\,$\times 7.6$\,kpc). We find several distinct tidal features
associated with \group, including clear evidence for an interaction with
NGC\,1047. No tidal debris is detected near \dftwo\ or \dffour.
}
\label{environ.fig}
\end{figure*}

With these assumptions we find an enclosed
mass within 7\,kpc of $M_{\rm TME}=0.4^{+1.2}_{-0.3}\times 10^8$\,\msun.
The total stellar mass is 
$M_{\rm stars}=(1.5 \pm 0.4)\times 10^8$\,\msun\ (see \S\,2),
and we conclude that we cannot reject the hypothesis that
there is no dark matter in this system.
Another way to phrase this is
that the intrinsic dispersion of $\sigma_{\rm
intr}=4.2^{+4.4}_{-2.2}$\,\kms\ is consistent with that
expected from the stars alone
($\sigma_{\rm stars}\approx 7.3^{+0.9}_{-1.1}$\,\kms; {Wolf} {et~al.} 2010).
We note that this result is independent of the precise distance to \dffour: 
as $\sigma_{\rm stars}\propto{}(M_{\rm stars}/R_{\rm e})^{0.5}$ it scales
with distance as $\sigma_{\rm stars}\propto(D^2/D)^{0.5}\propto{}D^{0.5}$.
For reference, the expected dispersion from an NFW halo of the expected
mass is $\sigma_{\rm halo}\approx 30$\,\kms\ 
({{\L}okas} \& {Mamon} 2001; {Behroozi} {et~al.} 2013a).
%We emphasize that this is not the dispersion in the center of the halo
%but at an average radius of $\langle R\rangle =3.5$\,kpc;
%Quantitative limits on the total dark matter mass require a
%more comprehensive analysis
%(see {Wasserman} {et~al.} 2018),
%but we note here that the most distant globular clusters almost reach
%the virial radius for halo masses of $\sim 10^8$\,\msun.

\section{Discussion}

In this {\em Letter} we have presented a doppelg\"anger
of the dark matter-deficient galaxy
\dftwo. \dffour\ 
is in the same group as \dftwo\
and has a similar size, luminosity, 
morphology, globular cluster population, and velocity
dispersion. 
The immediate implication is that \dftwo\ is not an isolated case but
that a class of such galaxies exists, and
given how little we know about galaxies in the UDG parameter space
it may well be that they are fairly common.

The discovery of \dffour\ does not bring us much closer to understanding how
such galaxies are formed, although it does effectively rule out
``tail of the distribution'' explanations for \dftwo.
Suggestions
that the true velocity dispersion is in the upper 10\,\%  of the
posterior distribution ({Martin} {et~al.} 2018), that the galaxy
could be an asymmetric thin disk seen exactly face-on
(see {van Dokkum} {et~al.} 2018b), is on a precisely-tuned
orbit
({Ogiya} 2018), or that it was formed in a carefully orchestrated sequence of
events (e.g., {Fensch} {et~al.} 2019), are far less likely now that there is a second system.

\begin{figure}[htbp]
  \begin{center}
  \includegraphics[width=0.9\linewidth]{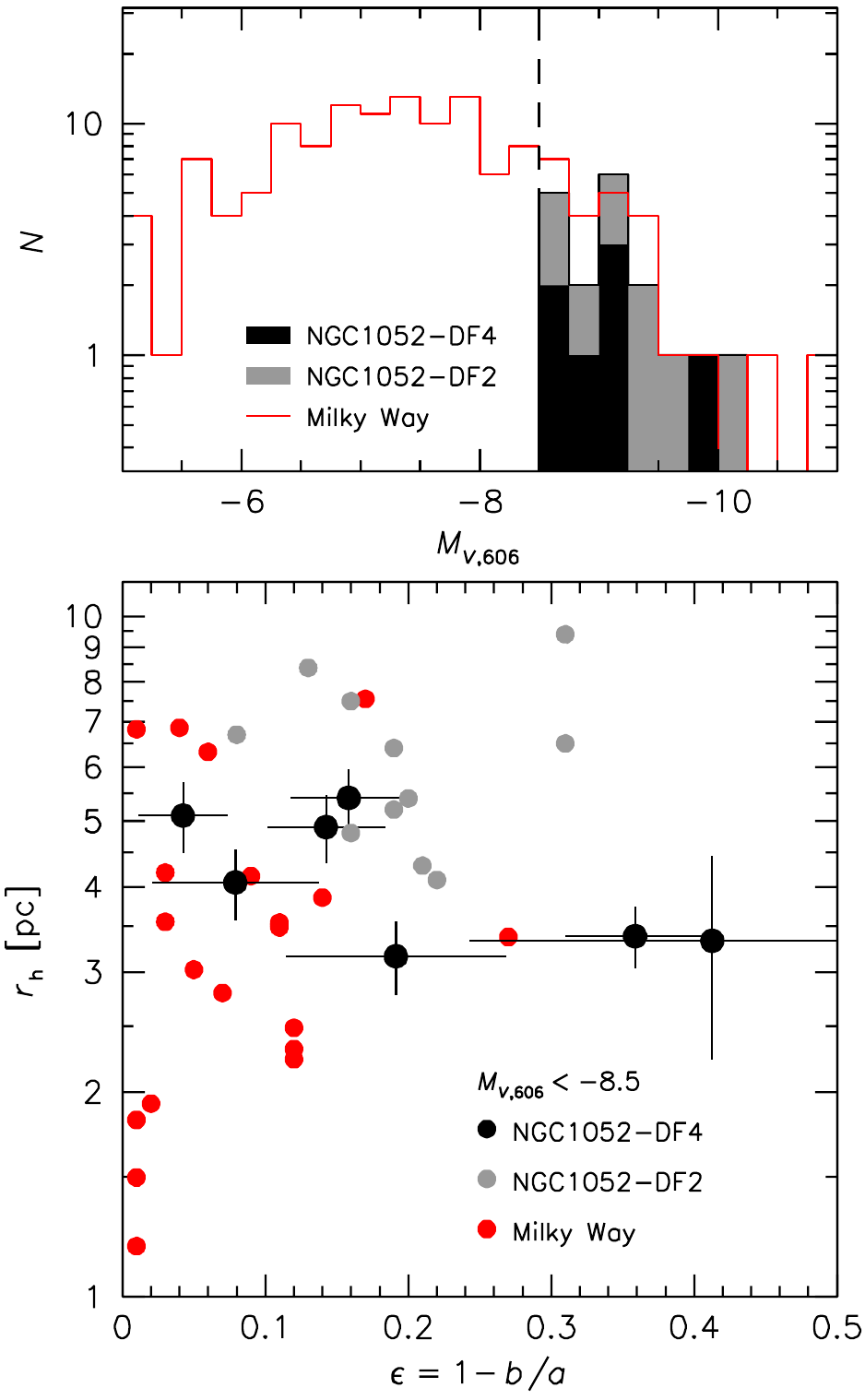}
  \end{center}
\vspace{-0.2cm}
    \caption{{\em Top panel:} Luminosity function of spectroscopically-confirmed globular
    clusters in \dftwo\ and \dffour, compared to that of the Milky Way. {\em Bottom panel:} half-light radii and ellipticities of the clusters. The confirmed objects in \dftwo\ and \dffour\ are similar to the most luminous globular clusters in the Milky Way ({Harris} 1996; {Harris}, {Harris}, \& {Alessi} 2013).
}
\label{structure.fig}
\end{figure}

One  pathway for creating 
dark matter-deficient 
galaxies is by forming them out of gas that was expelled from
a disk with a high baryon fraction, through a tidal interaction
(e.g., {Duc} \& {Mirabel} 1998; {Gentile} {et~al.} 2007). Although some
form of tidal origin is perhaps the most plausible explanation for
these objects,
\dftwo\ and
\dffour\ lack two of the key identifying features of ``classical" tidal
dwarf galaxies.
First, as the gas originated in a dense disk it
should be pre-enriched. Therefore, tidal
dwarfs should have a high metallicity for their mass ({Duc} \& {Mirabel} 1998),
and this is not the case for \dftwo\ ({van Dokkum} {et~al.} 2018c; {Fensch} {et~al.} 2019).
Second,  although we identify
several tidal features associated with NGC\,1052
in the Dragonfly imaging,
there is no evidence for
debris in the vicinity of \dftwo\ or \dffour, although this has
been reported around other old tidal dwarfs (see {Duc} {et~al.} 2014).

More broadly, the environment of \dffour\ does
not shed much light on its origins.
The galaxy is
highlighted in a wide field of view in Fig.\
\ref{environ.fig}.
As noted in \S\,4.1 the systemic velocity of \dffour\ is
almost identical to the average of the \group\ group galaxies. It
is at a projected distance of
$28\farcm 5$ (165\,kpc) from \group\ itself, a factor of two further than
\dftwo, and $26\arcmin$ (150\,kpc)
from the spiral galaxy NGC\,1042 (which is almost certainly also a
group member; see van Dokkum et al.\ 2019).
It is close (23\,kpc) in projection to NGC\,1035, which has a radial
velocity of $cz=1241$\,\kms. Given their velocity difference
of 204\,\kms\ 
it is unlikely that \dffour\ is a satellite of this low
mass disk galaxy ({Truong} {et~al.} 2017), which means that their 3D distance
in the group is probably much larger than their projected distance. Apart from the relatively
large systemic
velocity of \dftwo\ ($1805$\,\kms;
{Danieli} {et~al.} 2019) there is nothing obviously
``special'' about the two galaxies in relation to other group
members, or about the \group\ group when compared to other structures.

Although it is hazardous to draw
conclusions from such correlations, it seems likely that the
low dark matter mass in these galaxies is somehow
related to their unprecedented
globular cluster systems (see {van Dokkum} {et~al.} 2018c).
%\footnote{In {van Dokkum} {et~al.} (2018c)
%we speculated that \dftwo, and
%globular cluster-rich UDGs in general, may have started
%out as extremely dense systems that expanded due to intense
%supernova feedback during the formation of the clusters.}
The combined number of confirmed clusters with
$M_V<-8.5$ in \dftwo\ and \dffour\ is nearly the same as the
number of globular clusters in the Milky Way to that limit
(18 versus 23), despite the factor of 100 difference in stellar mass
between them (see Fig.\ \ref{structure.fig}).
The seven confirmed clusters in \dffour\ 
make up 3\,\% of its total luminosity, and 
the two most distant clusters by themselves
make up $\approx 70$\,\% of \dffour's luminosity at
$R>5$\,kpc.
%\footnote{We note that this is a lower limit: there
%may be clusters beyond the boundaries of the HST image,
%and even within the HST image our sample is incomplete.}
As shown in the bottom panel of Fig.\ \ref{structure.fig}
the half-light radii of the clusters (measured in the same way as
described in van Dokkum et al.\ 2018c) are somewhat smaller than
those in \dftwo, and similar to those of
luminous Milky Way clusters. The median size of the
seven clusters is $\langle r_{\rm h}\rangle =4.1$\,pc.

Looking ahead,  we can determine the stellar velocity dispersion 
of \dffour\
(see {Emsellem} {et~al.} 2019; {Danieli} {et~al.} 2019),  constrain its
dark matter mass (e.g., {Laporte} {et~al.} 2019; {Wasserman} {et~al.} 2018), and assess
the implications for alternative gravity
({van Dokkum} {et~al.} 2018b; {Famaey}, {McGaugh}, \& {Milgrom} 2018):
taken together, \dftwo\ and \dffour\ seem in tension with
recent predictions from Modified Newtonian Dynamics ({M{\"u}ller}, {Famaey}, \&  {Zhao} 2019).
Following the adage ``one is an exception but two is a population'' this
new object provides impetus for 
characterizing the properties of diffuse, dark matter-deficient galaxies 
as a class.
We are  performing wide field surveys with the Dragonfly Telephoto Array
to identify other candidates.

\acknowledgements{
We thank the anonymous referees for excellent comments, including the suggestion to
look for diffuse galaxy light in the sky background.
Support from STScI grants HST-GO-13682 and HST-GO-14644, as well as NSF grants
AST-1312376, AST-1613582, and AST-1515084 is gratefully acknowledged. 
%JMDK gratefully acknowledges funding from the German Research Foundation (DFG) in the form of an Emmy Noether Research Group (grant number KR4801/1-1) and from the European Research Council (ERC) under the European Union's Horizon 2020 research and innovation programme via the ERC Starting Grant MUSTANG (grant agreement number 714907).
AJR is a Research Corporation for Science Advancement Cottrell Scholar.
}

%% --------------------------------------------------------------------
%% Wed Jan 16 12:50:46 2019
%%   This file was generated automagically from the files
%%   df4.bbl and df4.tex using
%%     nat2jour.pl
%%   This file should accompany df4-aas.tex.
%% --------------------------------------------------------------------

\end{document}